\documentclass[12pt,preprint]{aastex}
\def\etal{et al.}

\begin{document}
\title{A Simple Analytical Formulation for Periodic Orbits in Binary Stars}

\author{Erick Nagel\altaffilmark{1} \& Barbara
  Pichardo\altaffilmark{1,2}} \altaffiltext{1}{Instituto de
  Astronom\'\i a, Universidad Nacional Aut\'onoma de M\'exico,
  Apdo. Postal 70-264, 04510, M\'exico, D.~F.,
  M\'exico.}\altaffiltext{2}{Institute for Theoretical Physics,
  University of Zurich, Winterthurerstrasse 190, Zurich 8057,
  Switzerland.}

\begin{abstract} 

An analytical approximation to periodic orbits in the circular restricted
three-body problem is provided. The formulation given in this work is
based in calculations known from classical mechanics, but with the
addition of the necessary terms to give a fairly good approximation
that we compare with simulations, resulting in a simple set of
analytical expressions that solve periodic orbits on discs of binary
systems without the need of solving the motion equations by numerical
integrations.

\end{abstract}


\section{INTRODUCTION}\label{sec:introd}

The majority of low-mass main-sequence stars seem to be grouped in
multiple systems preferentially binary (Duquennoy \& Mayor 1991,
Fischer \& Marcy 1992). These systems have attracted more attention
since the discovery that many T-Tauri and other pre-main-sequence
binary stars possess both circumstellar and circumbinary discs from
observations of excess radiation at infrared to millimeter waves and
direct images in radio (Rodr{\'i}guez \etal 1998; for a review see
Mathieu 1994, 2000). The improvement in the observational techniques
allows, on the other hand, the testing of theories on these
objects. Moreover, extrasolar planets have been found to orbit stars
with a stellar companion, e.g., 16 Cygni B, $\tau$ Bootis, and 55
$\rho$ Cancri (Butler \etal 1997; Cochran et al. 1997). All these
facts make the study of stellar discs in binary systems, as well as
the possibility of stable orbits on them that could be populated by
gas or particles, a key element for better understanding stellar and
planetary formation.

In particular the study of simple periodic orbits of a test particle
in the restricted three-body problem results a very good approximation
to the streamlines in an accretion disc in a very small pressure and
viscosity regime, or for proto-planetary systems and its debris
(Kuiper belt objects). Extensive theoretical work has been done in
this direction (Lubow \& Shu 1975; Paczy\'nski 1977; Papaloizou \&
Pringle 1977; Rudak \& Paczy\'nski 1981; Bonnell \& Bastien 1992, Bate
1997; Bate \& Bonnell 1997).

Several studies are directed to find the most simple and important
geometric characteristic of the discs which is the size of both
circumstellar and circumbinary discs, going from analytical
approximations: Eggleton (1983), who provides a simple analytical
approximation to the Roche lobes; Holman \& Wiegert (1999) radii in
planetary discs; Pichardo, Sparke \& Aguilar (2005, PSA05 hereafter)
radii in eccentric binary stars.

Based in the approximation studied in classical mechanics to periodic
orbits given by Moulton (1963), we provide in this work a fairly good
approximation by calculating the necessary terms, for the case of
circular binaries, to periodic orbits. This formulation gives not only
the radii of the circumstellar and circumbinary discs, but gives a
good description of any periodic orbit at a given radius for any of
the discs, in the form of an analytical approximation. In this manner,
the set of equations we provide can be used to find any periodic orbit
or its initial conditions to run it directly in simulations of the
three-body restricted problem, or as a good approximation to initial
conditions in the eccentric case as well, from the point of view of
particle or hydrodynamical simulations, without the need of solving
the restricted three body problem numerically.

The low viscosity regime described by the periodic orbits
representation, is important in astrophysics since it gives an idea of
how bodies would respond only influenced by the effects of the
potential exerted by the binaries, and they could permit to link the
periodic orbits to physical unknown characteristics of the discs like
viscosity that needs to be artificially introduced to hydrodynamical
codes. It also might permit the calculation of physical
characteristics like dissipation rates, etc. Studies like these result
easier given in the form of approximated analytical expressions,
instead of the usual numerical approximations to the periodic orbits
since an analytical formulation is much faster to add to any
hydrodynamical or particle code that requires the characteristics of a
given family of periodic orbits, or simply their initial conditions in
a more precise form than assuming Keplerian discs until the
approximation fails. The periodic orbits show regions where the orbits
are compressed (or decompressed), these regions could trigger density
fluctuations maybe able to drive material to form important
agglomerations in the discs that, depending on their positions on the
disc and the density, could give origin to planets. An analytical
approximation with a given density law for the discs would allow these
kind of studies to be much faster and easier.

In Section \ref{sec:method}, we show the strategy and equations used
to find the approximation to the periodic orbits in circumstellar
discs. In Section \ref{sec:orbits2}, we present the equations
approximating the periodic orbits in circumbinary discs, and an
analysis of stability in Section \ref{sec:stability}. A comparison of
the application of the formulation and numerically calculated periodic
orbits is given in Section \ref{sec:comparison}.  Conclusions are
presented in Section \ref{sec:conclusions}.

\section{Methodology}\label{sec:method}

In general, orbits around binary systems are calculated by approximate
methods. Numerical calculations are the most extended. In this work we
are going to use other known method to search for an analytical
approximation that will provide us of solutions for any orbit in
circumbinary or circumstellar discs of the circular problem. The
method is based on a perturbative analysis. In this case we choose
some terms in the expanded equations of motion, which are relevant to
the solution of the problem. The original equations are approximated
with a set of equations that, in many cases, have an analytical
solution. Comparing with the numerical approach, this one has the
advantage that the expressions are completely analytical and
simple. We compare our calculations with numerical studies in order to
pick as many terms in the expansion as we need to obtain the solution.
The method can be used as a way to find a quick answer and for studies
in which an analytic expression makes the problem manageable.

\subsection{Periodic Orbits in Circumstellar Discs}
\label{sec:orbits1}

The analysis is restricted to the orbital plane, in polar coordinates
the equations of motion are given by,

\begin{eqnarray}
    \ddot{r}-r\dot{\psi}^{2} &=& F_{r}, \nonumber \\
    r\ddot{\psi}+2\dot{r}\dot{\psi} &=& F_{\psi},
    \label{eq:motion}
\end{eqnarray}

\noindent where the origin of the system is located at the position of
the star with mass $M_{1}$ (hereafter we call this star $S_{1}$). It
is important noticing that $S_{1}$ can represent either the primary or
the secondary star.  Here, $r$ and $\psi$ are the radial and angular
coordinates respectively, and the forces along this directions
($F_{r}$,$F_{\psi}$) have the form

\begin{eqnarray}
    F_{r} &=& -{ GM_{1}\over r^{2} } + GM_{2}{\partial \Phi_p\over \partial r},
           \nonumber \\
    F_{\psi} &=& {GM_{2}\over r}{\partial \Phi_p\over \partial \psi}, \nonumber
\end{eqnarray}

\noindent with

\begin{equation}
   \Phi_p={1\over (R^{2}-2rR\cos (\psi-\Psi)+r^{2})^{1/2}}-
          {r\cos (\psi-\Psi)\over R^{2}},
    \label{eq:disturb_potential}
\end{equation}

\noindent that represents the perturbing potential (divided by $GM_2$)
due to the star with mass $M_{2}$ (hereafter, star $S_{2}$). $\Phi_p$
could be also expressed in terms of the Legendre Polynomials (Murray
\& Dermott, 1999). Here $R$ and $\Psi$ are the coordinates of $S_{2}$
and $r$ and $\psi$ locate the particle that is perturbed, $P$.

The non-perturbed state corresponds to the case with $\Phi_p=0$. Thus, the
non-perturbed orbits are conics around $S_{1}$, which are perturbed by
the presence of the other mass. The set of equations (\ref{eq:motion})
cannot be solved analytically, then we expand the perturbation given
in equation (\ref{eq:disturb_potential}) using $r/R$ as a small
parameter. This parameter at some points in an orbit that begins around a 
resonance is large, so this approach is not valid. This expansion improves in 
precision the closer the
particle is to $S_{1}$. Thus, equations (\ref{eq:motion}) can be
written as,

\begin{eqnarray}
\ddot{r}-r\dot{\psi}^{2}+{GM_{1}\over r^{2}}&=&{GM_{2}\over 2}{r\over R^{3}}
        \left[(1+3\cos 2(\psi-\Psi)) + {3\over 4}
{r\over R} (3 \cos(\psi-\Psi)+5 \cos3(\psi-\Psi))+\right. \nonumber\\
&+& \left. \left({r\over R}\right)^2\left({9\over 8}+{5\over 2}\cos 2(\psi-\Psi)+{35\over8}\cos 4(\psi-\Psi)\right)+ ...\right], \nonumber \\
r\ddot{\psi}+2\dot{r}\dot{\psi} &=& -{GM_{2}\over 2}{r\over R^{3}}
\left[(3\sin 2(\psi-\Psi)) + {3\over 4}
{r\over R} (\sin(\psi-\Psi)+5\sin3(\psi-\Psi))+\right. \nonumber\\
&+& \left. {5\over 4}\left({r\over R}\right)^2\left(\sin 2(\psi-\Psi)+{7\over2}\sin 4(\psi-\Psi)\right)+ ...\right]\; ,
\label{eq:expanded_eqns}
\end{eqnarray}

\noindent which is equivalent to the expansion (6.22) in 
Murray \& Dermott (1999).

In this analysis the star $S_{2}$ orbits $S_{1}$ describing a circular
orbit. We consider the mass of the discs negligible compared with the
mass of the binary system. Thus, the orbit of $S_{2}$ is given by

\begin{eqnarray}
    R &=& a, \nonumber \\
    \Psi &=& \sqrt{G(M_{1}+M_{2})\over a^3} (t-t_{0})=\Omega(t-t_{0}),
    \label{eq:orbit_s2}
\end{eqnarray}

\noindent where $a$ represents the fixed distance between the stars,
and $\Omega$ is the angular velocity of $S_{2}$ from the third law of
Kepler. At $t=t_{0}$ the star is in the reference line, which for a
circular orbit can be any radial line that begins on $S_{1}$. The
non-perturbed trajectory of the particle $P$, is a Keplerian orbit
around $S_{1}$, in this manner the polar coordinates are expressed,

\begin{eqnarray}
    r &=& a_{np}, \nonumber \\
    \psi &=& \sqrt{GM_{1}\over a_{np}^3} (t-t_{0})=\omega(t-t_{0}),
    \label{eq:orbit_p}
\end{eqnarray}

\noindent where $a_{np}$ and $\omega$ have the same meaning of its counterpart
in equation \ref{eq:orbit_s2}, but relating $P$ and $S_{1}$.

The perturbed position of the particle $P$ can be written,

\begin{equation}
    r=a_{np}(1+r_p), 
    \label{eq:disturbed_r}
\end{equation}

\begin{equation}
    \psi=\omega(t-t_{0})+\psi_p,
    \label{eq:disturbed_v}
\end{equation}

\noindent where the terms $(a_{np}r_p)$ and $\psi_p$ are the
corrections in the trajectory due to the presence of $S_{2}$. We
define now the parameters $m_o$ and $\tau$ as follows,

\begin{equation}
    m_{o}={\Omega\over \omega-\Omega},
    \label{eq:m}
\end{equation}
\begin{equation}
    \tau=(\omega-\Omega)(t-t_{0}),
    \label{eq:tau}
\end{equation}

\noindent here $m_{o}$ is a small parameter for $r\ll R$. This
condition is required for the expansion of the equations of motion
(equations \ref{eq:expanded_eqns}) to be useful.  $\tau$ is the angle
that locates the particle $P$, measured from the line connecting the
stars, for the non-perturbed orbit. The first step is to replace $t$
for $\tau$ in equation (\ref{eq:expanded_eqns}), using equation
(\ref{eq:tau}).  Equations (\ref{eq:disturbed_r} and
\ref{eq:disturbed_v}) are substituted in equations
(\ref{eq:expanded_eqns}) taking into account that $r_p$ and $\psi_p$
depend on $\tau$. The resulting expressions are finally divided by
$a_{np}({\Omega\over m_o})^2$. The mass is given in units of
$M_{1}+M_{2}$, and $M_{1}+M_{2}=1$, then,

\begin{eqnarray}
   \ddot{r_p}-(1+r_p)(1+m_{o}+\dot\psi_p)^2+{(1+m_{o})^2\over (1+r_p)^2}&=\nonumber\\
     {m_{o}^2M_{2}\over 2}(1+r_p)[(1+3\cos 2(\tau+\psi_p)) &\nonumber\\
+{3\over4} M_p m_0 (1+r_p)(3\cos (\tau+\psi_p)+5\cos 3(\tau+\psi_p))&\nonumber\\
+M_p^2 m_0^2 (1+r_p)^2 ({9\over8}+{5\over2}\cos2(\tau+\psi_p)+{35\over8}\cos 4(\tau+\psi_p))&\nonumber\\
+...]&,
    \label{eq:motion_rho}
\end{eqnarray}

\begin{eqnarray}
    (1+r_p)\ddot\psi_p + 2\dot{r_p}(1+m_{o}+\dot\psi_p)&=\nonumber\\
     -{m_o^2M_2\over 2}(1+r_p)[3\sin 2(\tau+\psi_p) &\nonumber\\
+{3\over4} M_p m_0 (1+r_p)(\sin (\tau+\psi_p)+5\sin 3(\tau+\psi_p))&\nonumber\\
+{5\over4}M_p^2 m_0^2 (1+r_p)^2 (\sin2(\tau+\psi_p)+{7\over2}\sin 4(\tau+\psi_p))&\nonumber\\
+...]&  \; .
    \label{eq:motion_psip}
\end{eqnarray}

Equations (\ref{eq:motion_rho} and \ref{eq:motion_psip}) are
dimensionless. They will be solved for $r_p$ and $\psi_p$ in terms of
$M_{2}$ ($M_{1}=1-M_{2}$), $\tau$ and $m_{o}$. The value for $M_{2}$,
characterizes the stellar masses of the system, $\tau$ defines the
angular position of the point in the original non-perturbed orbit, and
$m_{o}$ depends on the radial position of the non-perturbed trajectory
$a_{np}$, in units of the separation between the stars, $a$.  An analytic
solution is found if an expansion for $r_p$ and $\psi_p$ in powers of
$m_{o}$ is made as follows,

\begin{equation}
    r_p=\Sigma_{j=2}^{\infty}r_{p_j}(M_{2},\tau)m_{o}^{j},
    \label{eq:expansion_rho}
\end{equation}
\begin{equation}
    \psi_p=\Sigma_{j=2}^{\infty}\psi_{p_j}(M_{2},\tau)m_{o}^{j}.
    \label{eq:expansion_psip}
\end{equation}

Specifically, for periodic orbits, it is necessary to fulfill the next
conditions,

\begin{eqnarray}
    r_{p_j}(\tau+2\pi) &=& r_{p_j}(\tau), \nonumber \\
    \psi_{p_j}(\tau+2\pi) &=& \psi_{p_j}(\tau).
    \label{eq:periodic_conditions}
\end{eqnarray}

An useful assumption is that the angular position of the orbit, from
the line that connects the two stars, is not perturbed by $S_{2}$, a
fact necessarily true, which can be demonstrated using symmetry
arguments. We restrict the perturbed orbits to be symmetric with
respect to the line joining the stars by,

\begin{equation}
    \psi_{p_j}(0)=0,
    \label{eq:condition_tau0}
\end{equation}

\noindent this condition is general enough since there are no
restrictions provided by observations about the orientation of the
symmetry axis in the plane of the discs with respect to the line
joining the stars.

Equations (\ref{eq:expansion_rho} and \ref{eq:expansion_psip}) are
substituted in equations (\ref{eq:motion_rho} and
\ref{eq:motion_psip}) and expanded in powers of $m_{o}$, ending with a
polynomial in $m_{o}$ equal to zero. The only solution for such an
equation is that every coefficient is zero. Each coefficient
associated to a given power, is a second order differential
equation. The solutions of the equations for $m_{o}^k$ with $k=2,3,4$,
are solved imposing the conditions given in equations
(\ref{eq:periodic_conditions} and \ref{eq:condition_tau0}) and
are given by

\begin{equation}
    r_{p_2}=-{M_{2}\over 6}+{15\over16}HM_2\cos \tau - M_{2}\cos 2\tau,
    \label{eq:rho2}
\end{equation}
\begin{equation}
    \psi_{p_2}=-{15\over8}HM_2\sin\tau + {11M_{2}\over 8}\sin 2\tau,
    \label{eq:psip2}
\end{equation}

\begin{equation}
    r_{p_3}={M_{2}\over 3}+\bigl{(}{135\over 32}HM_{2}^{2}-{3\over 32}HM_{2}
       \bigr{)}\cos \tau - {7\over 6}M_{2}\cos 2\tau - 
	    {25\over 64}HM_{2}\cos 3\tau,
    \label{eq:rho3}
\end{equation}
\begin{equation}
    \psi_{p_3}=-\bigl{(}{135\over 16}HM_{2}^{2}+{21\over 16}HM_{2}\bigr{)}\sin
       \tau + {13\over 6}M_{2}\sin 2\tau + {15\over 32}HM_{2}\sin 3\tau,
    \label{eq:psip3}
\end{equation}
\begin{eqnarray}
    r_{p_4} &=& {331\over 288}M_{2}^{2}-{1\over 2}M_{2}+{225\over 512}
       H^{2}M_{2}^{2}-{3\over 16}H^{2}M_{2}+q_{4}\cos\tau + \nonumber \\
       & & \bigl{(}{2\over 3}M_{2}^{2}-{10\over 9}M_{2}-{225\over 512}
       H^{2}M_{2}^{2}-{5\over 8}H^{2}M_{2}\bigr{)}\cos 2\tau + \nonumber \\
       & & \bigl{(}{255\over 256}HM_{2}^{2}-{65\over 256}HM_{2}\bigr{)}
       \cos 3\tau
       -\bigl{(}{3\over 8}M_{2}^{2}+{7\over 32}H^{2}M_{2}\bigr{)}\cos 4\tau,
       \label{eq:rho4}
\end{eqnarray}
\begin{eqnarray}
    \psi_{p_4} &=& -\bigl{(}2q_{4}+{515\over 64}HM_{2}^{2}-{3\over 16}HM_{2}
      \bigr{)}\sin\tau + \bigl{(}-{1\over 2}M_{2}^{2}+{41\over 18}M_{2}+
       {1125\over 1024}H^{2}M_{2}^{2}+{25\over 32}H^{2}M_{2}\bigl{)}\sin 2\tau
       \nonumber \\
       & & +\bigl{(}-{255\over 128}HM_{2}^{2}+{55\over 128}HM_{2}\bigr{)}
       \sin 3\tau
       +\bigl{(}{201\over 256}M_{2}^{2}+{63\over 256}H^{2}M_{2}\bigr{)}
       \sin 4\tau,
       \label{eq:psip4}
\end{eqnarray}

with

\begin{equation}
    q_{4}={3\over 64}M_{2}H-{1727\over 256}M_{2}^{2}H+{1215\over 64}M_{2}^{3}H
          +{105\over 128}M_{2}H^{3}.
\end{equation}

\noindent It is worth noticing that for each order (in $m_0^k$) we
obtain a system of two differential equations that produce two
integration constants that can be calculated by solving the set of
equations of the next higher order, by imposing the conditions in
equations \ref{eq:periodic_conditions} and \ref{eq:condition_tau0}.

The parameter $H$ comes from the approximation $a_{np}=m_{o}H$, where
we will consider $H$ constant, justified by the fact that $H$ depends
weakly on $m_{o}$ (Moulton 1963).

Using equations (\ref{eq:rho2}, \ref{eq:psip2}, \ref{eq:rho3},
\ref{eq:psip3}, \ref{eq:rho4} and \ref{eq:psip4}), plus equations
(\ref{eq:expansion_rho} and \ref{eq:expansion_psip}) in equations
(\ref{eq:disturbed_r} and \ref{eq:disturbed_v}) will let us find the
perturbed trajectory of the particle $P$. Since $a_{np}$ is
proportional to $m_{o}$, the radial corrections are of order
$m_{o}^{5}$ and the angular corrections are of order
$m_{o}^{4}$. Thus, equations (\ref{eq:disturbed_r} and
\ref{eq:disturbed_v}) represent coordinates depending on time, which
means velocity and accelerations can be obtained by direct
differentiation.  It is worth noticing that the orbits are given in
the reference system where the stars are fixed in the $X$ axis. An
example of the orbits and rotation curves produced by this model are
given in Section \ref{sec:comparison}.

\subsection{The Radial Extension of Circumstellar Discs}
\label{sec:size1}

To calculate the total radial size of discs we will assume we are in
the regime of low pressure gas. The permitted orbits for gas to settle
down there, are all those that do not intersect themselves or any
other (Paczy\'nski 1977, PSA05). In this paper we will use the same
approximation to find discs radii except that the intersections are
found analytically (other analytical approximation to the radial
extension for circumstellar discs for any eccentricity can be found in
PSA05). First, we choose a given $M_{2}$, to fix the binary system. A
physical intersection occurs when

\begin{equation}
    r_{2}-r_{1}=0,
    \label{eq:inters_r}
\end{equation}
\begin{equation}
    \psi_{2}-\psi_{1}=0,
    \label{eq:inters_v}
\end{equation}

\noindent where $r_{i}$ is the radius of the orbit given by equation
(\ref{eq:disturbed_r}) and $\psi_{i}$ spans the $2\pi$-angular-range of
the same orbit. The subindex $i$ takes the values $i={1,2}$ for inner
and outer consecutive orbits, respectively. We now look for a value
for $a_{np}$ and $\tau$ that simultaneously satisfy equations
(\ref{eq:inters_r} and \ref{eq:inters_v}). Note that equation
(\ref{eq:inters_r}) can be expanded in a series of Taylor in the
variable $a_{np}$. Due to the assumption that we are interested in
infinitesimally close orbits, a linear approximation is good
enough. In this manner, equation (\ref{eq:inters_r}) can be
transformed to

\begin{equation}
    {dr\over da_{np}}=0,
    \label{eq:intersec_r}
\end{equation}

\noindent which is a function of $M_{2}$,$a_{np}$ and $\tau$. Also, using
equation (\ref{eq:disturbed_v}), the equation (\ref{eq:inters_v}) can
be written as

\begin{equation}
    \psi_{p,2}-\psi_{p,1}=0.
    \label{eq:intersec_v}
\end{equation}

\noindent From equations (\ref{eq:psip2}, \ref{eq:psip3} and
\ref{eq:psip4}) we directly identify that all the terms are
proportional to $\sin (k\tau)$, which results proportional to
$\sin\tau$. Thus, $\sin\tau$ can be factorized in equation
(\ref{eq:intersec_v}) giving the solutions $\tau = 0$ and $\tau =
\pi$. Equation (\ref{eq:intersec_v}) has other solutions but none
allows to find a root for equation (\ref{eq:intersec_r}). Taking
($\tau=0$,$\tau=\pi$) and using any of these options in equation
(\ref{eq:intersec_r}), we end up with an equation for $a_{np}$. The
solution of equation (\ref{eq:intersec_r}) using $\tau=0,\pi$ gives
two values of $a_{np}$, which correspond to the intersections.  We
take the intersections at the smallest radius (that represent the
maximum radii where gas particles would be able to settle down), thus
the second intersection at larger radius are not useful in this
analysis. In both cases the intersecting orbits are tangent to each
other, and contiguous orbits between them intersect at an
angle. Because we do not consider this kind of orbits as part of the
disc, the disc naturally ends up in the orbits with smaller $a_{np}$.

The minimum value for $a_{np}$ comes from $\tau=\pi$, this is, the
intersection appears at the opposite side of the star $S_{2}$. In
Table \ref{table.csdiscs}, the solution of equation
(\ref{eq:intersec_r}) for different values of $M_{1}$, with
$\tau=\pi$, is given in the column named $a_{np,\tau=\pi}$. The
following two columns ($r(\tau=0)$,$r(\tau=\pi)$) give the radial
positions of the innermost intersecting orbit. The average of these
values is given in the column $<r>$, the next column gives the same
average but from PSA05, and the last column is the approximation to
the Roche Lobe radius by Eggleton (1983) given by,

\begin{eqnarray}
\frac{R_i}{a}
\approx  \frac{R_{i\rm (Egg)}}{a}
=\frac{0.49 q_i^{2/3}}{0.6q_i^{2/3}+ln({1+q_i^{1/3}})}
\, ,~ {\rm where} ~~ \\   
q_1=\frac{1-M_2}{M_2} \,, \\
q_2=\frac{M_2}{1-M_2} \,, \\ 
\label{eq.roche}
\end {eqnarray}

\noindent where $i$ refers to either star $S_i$ with $i=1,2$.

\begin{table*}
 \centering
% \begin{minipage}{140mm}
  \caption{Sizes of the circumstellar discs.}
  \begin{tabular}{@{}ccccccc@{}}
%    Name     &            & \multicolumn{4}{c}{Flux density (Jy)%
   $M_{1}$ & $a_{np,\tau=\pi}$ & $r(\tau=0)$ & $r(\tau=\pi)$ & $<r>$
     & $<r>_{PSA}$& $R_{RL}$\\
%        &  &  &  &  &  & group & (d) & curve \\
%        &  &  &  &  &  &       &     & type  \\[10pt]
 0.1 & 0.1628 & 0.1641 & 0.1326 & 0.1484 & 0.125 & 0.213\\
 0.2 & 0.2077 & 0.2113 & 0.1690 & 0.1901 & 0.162 & 0.268\\
 0.3 & 0.2445 & 0.2486 & 0.1989 & 0.2237 & 0.195 & 0.308\\
 0.4 & 0.2795 & 0.2825 & 0.2276 & 0.2550 & 0.228 & 0.344\\
 0.5 & 0.3155 & 0.3155 & 0.2573 & 0.2864 & 0.257 & 0.379\\
 0.6 & 0.3543 & 0.3489 & 0.2901 & 0.3195 & 0.317 & 0.414\\
 0.7 & 0.3982 & 0.3841 & 0.3285 & 0.3563 & 0.350 & 0.454\\
 0.8 & 0.4507 & 0.4236 & 0.3766 & 0.4001 & 0.387 & 0.501\\
 0.9 & 0.5213 & 0.4774 & 0.4466 & 0.4620 & 0.426 & 0.570
\label{table.csdiscs}
\end{tabular}
%\end{minipage}
\end{table*}

From Table \ref{table.csdiscs}, $<r>$ and $<r>_{PSA}$ (for particular
masses of the stars, $M_{1}$ and $M_{2}=1-M_{1}$) coincides within a
$20\%$ of error for $M_{1} \leq 0.5$, decreasing to less than $10\%$
for $M_{1} > 0.5$. As expected the last non-intersecting orbit is
contained inside the Roche lobe, as we can see in the Figure
\ref{fig.compEB_mosaico_q0.2} for $M_{2}=0.2$. A good approximation
for the Roche lobe is given using the approximation of Eggleton
(1983).

\section{Orbits in Circumbinary Discs}
\label{sec:orbits2}

The method described in Section \ref{sec:orbits1} is followed in this
case with few differences. First of all, the origin is now on the
center of mass of the system. The perturbing potential $\Phi_p$, takes
the form,

\begin{equation}
   \Phi_p={GM_{1}\over r}\Bigl{[}1-{2R_{1}\over r}\cos (\psi-\psi_{1})+\bigl{(}
       {R_{1}\over r}\bigr{)}^{2}\Bigr{]}^{-1/2} +
      {GM_{2}\over r}\Bigl{[}1+{2R_{2}\over r}\cos (\psi-\psi_{1})+\bigl{(}
       {R_{2}\over r}\bigr{)}^{2}\Bigr{]}^{-1/2},
    \label{eq:disturb_potential2}
\end{equation}

\noindent where $R_{1}=M_{2}R$, $R_{2}=M_{1}R$ are the distance to the
star $S_{1}$ and $S_{2}$ from the origin, respectively. It is
important to mention that $S_{2}$ is located to the left of the center
of mass, and that it can represent either the secondary or the primary
star. $a$ is the distance between the stars. $r$ and $\psi-\psi_{1}$
are the coordinates for the particle $P$ that trace the circumbinary
orbit, where the angle is measured from the radial vector which aims
to $S_{1}$. In this case, equation (\ref{eq:motion}) can be used with
the new definitions as follows,

\begin{eqnarray}
    F_{r} &=& {\partial \Phi_p\over \partial r}, \\
    F_{\psi} &=& {1\over r}{\partial \Phi_p\over \partial \psi}.
\end{eqnarray}

Thus, an expansion of equation (\ref{eq:motion}) for ${R_{1}\over
r}\ll 1$ and ${R_{2}\over r}\ll 1$ can be developed,

\begin{eqnarray}
    \ddot{r}-r\dot{\psi}^{2}+{G(M_{1}+M_{2})\over r^{2}} &=& -GM_p
   {a^{2}\over r^{4}}\Bigl{\{}{3\over 4}[1+3\cos 2(\psi-\psi_{1})]+\nonumber \\
    & & {R_{1}-R_{2}\over 2r}[3\cos (\psi-\psi_{1})+5\cos 3(\psi-\psi_{1})] 
	 +...\Bigr{\}},\nonumber \\
     r\ddot{\psi}+2\dot{r}\dot{\psi} &=& -GM_p{a^{2}\over r^{4}}
         \Bigl{\{}{3\over 2}\sin 2(\psi-\psi_{1}) + \nonumber \\
     & & {3(R_{1}-R_{2})\over 8r}[\sin (\psi-\psi_{1})+5\sin 3(\psi-\psi_{1})] 
	 + ... \Bigr{\}},
     \label{eq:expanded_eqns2}	 
\end{eqnarray}

\noindent where $M_p=M_1 M_2 = M_2 (1-M_2)$.

The third term on the left side of the first equation represents the
contribution to the force, in the case that all the stellar mass is
concentrated in the origin. The right side of both equations are the
perturbing force due to the mass distribution between the stellar
components. If the right side of equation (\ref{eq:expanded_eqns2}) is
equal to zero, we find the non-perturbed orbit in the form,

\begin{eqnarray}
    r &=& a_{np}, \\
    \psi &=& {\sqrt{G(M_{1}+M_{2})\over a_{np}^{3}}}(t-t_{0})=\omega(t-t_{0}).
    \label{eq:orbit_p2}
\end{eqnarray}

Note that the definition of $\omega$ differs from the circumstellar case in
the mathematical expression, however the meaning is the same,
non-perturbed angular velocity of the particle $P$. Again, the
perturbed position of $P$ is given by equations (\ref{eq:disturbed_r}
and \ref{eq:disturbed_v}). Equation (\ref{eq:tau}) changes to

\begin{equation}
    \tau=(\Omega-\omega)(t-t_{0}),
    \label{eq:tau2}
\end{equation}

\noindent where $\Omega={\sqrt{G(M_{1}+M_{2})\over a^{3}}}$ is the
angular velocity of the binary system, then $\tau$ represents the
angle between $S_{1}$ and $P$.

An useful parameter is given by

\begin{equation}
    \mu={\omega\over\Omega}=\bigl{(}{a\over a_{np}}\bigr{)}^{3/2},
    \label{eq:mu}
\end{equation}

\noindent where $\mu$ is the parameter used to expand equation
(\ref{eq:expanded_eqns2}), then $\mu$ is small as we are only taking
into account circumbinary orbits far away from the stars. Changing
$t\rightarrow \tau$ (equation \ref{eq:tau2}) and expanding equation
(\ref{eq:expanded_eqns2}) in a power series of $\mu^{1/3}$, we can
write the equations of motion as

\begin{eqnarray}
   & & \ddot{r_p}-(1+r_p)\bigl{(}{\mu\over 1-\mu}+\dot{\psi_p}\bigr{)}^{2}+
         {\mu^{2}\over (1-\mu)^{2}}{1\over (1+r_p)^{2}} = \nonumber \\
   & &     -{M_p\over (1-\mu)^{2}}{\mu^{10/3}\over (1+r_p)^{4}}\Bigl{\{}
	 {3\over 4}[1+3\cos 2(\tau+\psi_p)] + \nonumber \\ 
   & &   {(2M_{2}-1)\mu^{2\over 3}\over
	 2(1+r_p)}[3\cos (\tau+\psi_p)+5\cos 3(\tau+\psi_p)]+...\Bigr{\}},
    \label{eq:motion_rho2}
\end{eqnarray}

\begin{eqnarray}
   & & (1+r_p)\ddot{\psi_p}+2\dot{r_p}\bigl{(}{\mu\over 1-\mu}+\dot{\psi_p}
     \bigr{)} = \nonumber \\
   & &  -{M_p\over (1-\mu)^{2}}{\mu^{10/3}\over (1+r_p)^{4}}\Bigl{\{}
	 {3\over 2}\sin 2(\tau+\psi_p)+ \nonumber \\
   & &   {3\over 8}{(2M_{2}-1)\mu^{2\over 3}
	 \over(1+r_p)}[\sin (\tau+\psi_p)+5\sin 3(\tau+\psi_p)]+...\Bigr{\}}.
    \label{eq:motion_psip2}
\end{eqnarray}

In the same way as in Section \ref{sec:orbits1}, differential
equations are extracted at different orders if we expand $r_p$ and
$\psi_p$ as follows,

\begin{equation}
    r_p=\Sigma_{i=4}^{\infty}r_{p_i}(M_{2},\tau)\mu^{i/3},
    \label{eq:expansion_rho2}
\end{equation}
\begin{equation}
    \psi_p=\Sigma_{i=4}^{\infty}\psi_{p_i}(M_{2},\tau)\mu^{i/3}.
    \label{eq:expansion_psip2}
\end{equation}
 
The perturbed trajectories must be closed, then equations
(\ref{eq:periodic_conditions}) are imposed. Also, the fact that the
orbit is symmetric with respect to the line joining the two stars,
allows to find the following condition,

\begin{equation}
    \dot{r_p}_{i}(\tau=0)=0.
    \label{eq:condition_rho2}
\end{equation}

Substitution of equations (\ref{eq:expansion_rho2} and 
\ref{eq:expansion_psip2}) in equations (\ref{eq:motion_rho2} and 
\ref{eq:motion_psip2}) gives a couple of polynomials in $\mu^{i/3}$. A 
solution can be found if every coefficient of the $\mu^{i/3}$-terms is zero.
Using conditions given in equations (\ref{eq:condition_rho2},
\ref{eq:periodic_conditions}) allows directly to find
the set of solutions.

Moulton (1963) describes the method and gives explicitly the
expressions for $r_{p_i}$ and $\psi_{p_i}$ for $i\leq 15$. The
integration constants with the full expressions are provided in the
next equations

\begin{eqnarray}
    {r_p}_{4} &=& {1 \over 4}M_p, \nonumber \\
    {r_p}_{5} &=& 0, \nonumber \\
    {r_p}_{6} &=& 0, \nonumber \\
    {r_p}_{7} &=& 0, \nonumber \\
    {r_p}_{8} &=& {3 \over 64}M_p(5(1-M_2)^2-9M_p+5M_{2}^2), \nonumber \\
    {r_p}_{9} &=& 0, \nonumber \\
    {r_p}_{10} &=& {9 \over 16}M_p^2\cos 2\tau, \nonumber \\
    {r_p}_{11} &=& 0, \nonumber \\
    {r_p}_{12} &=& {1 \over 768}M_p(175(1-M_2)^{4}-535(1-M_2)^{3}M_{2}+711(1-M_2)^{2}M_{2}^{2}-535(1-M_2)M_{2}^{3}+175M_{2}^{4}) 
              \nonumber \\
              & & -{M_p\over 2}(1-2M_{2})(3\cos \tau+
	       {5 \over 9}\cos 3\tau), \nonumber \\
    {r_p}_{13} &=& {3 \over 4}M_p\cos 2\tau, \nonumber \\
    {r_p}_{14} &=& -{39 \over 32}M_p^2+{M_p\over64}(25(1-M_2)^2-61M_p+25M_2^2)\cos2\tau+ \nonumber \\
&&{175\over 1024} M_p((1-M_2)^2-M_p+M_2^2)\cos4\tau, \nonumber \\
    {r_p}_{15} &=& -{9 \over 4}M_p(1-2M_{2})(\cos \tau
              +{5 \over 27}\cos 3\tau),
\label{rpsolution}  
\end{eqnarray}

\noindent and for the angle we have,

\begin{eqnarray}
    {\psi_p}_{4} &=& 0, \nonumber \\
    {\psi_p}_{5} &=& 0, \nonumber \\
    {\psi_p}_{6} &=& 0, \nonumber \\
    {\psi_p}_{7} &=& 0, \nonumber \\
    {\psi_p}_{8} &=& 0, \nonumber \\
    {\psi_p}_{9} &=& 0, \nonumber \\
    {\psi_p}_{10} &=& {3 \over 8}M_p\sin 2\tau, \nonumber \\
    {\psi_p}_{11} &=& 0, \nonumber \\
    {\psi_p}_{12} &=& -{3 \over 8}M_p(1-2M_{2})(\sin\tau+{5 \over 9}\sin 3\tau), \nonumber \\
    {\psi_p}_{13} &=& {3 \over 16}M_p\sin 2\tau, \nonumber \\
    {\psi_p}_{14} &=& {5 \over 32}M_p((1-M_2)^2-4M_p+M_2^2)\sin2\tau+ {35\over 256} M_p((1-M_2)^2-M_p+M_2^2)\sin4\tau, \nonumber \\
    {\psi_p}_{15} &=& {1 \over 4}M_p(1-2M_{2})(9\sin\tau-
                {25 \over 27}\sin 3\tau),
\label{psipsolution}  
\end{eqnarray}

\noindent where the integration constant for the radial equation
\ref{eq:motion_rho2} is calculated using the solution for six orders
ahead. While for the angular equation \ref{eq:motion_psip2} the
integration constant is obtained from the solution for ten orders
ahead.

It is worth to mention that the approximation to circumbinary periodic
orbits given by Moulton is far from the solution as we have compared
with the numerical solution, thus we have developed all the necessary
terms in the expansion of the potential to reach a fairly good
approximation to the numerical solution of the problem.

In this way, an analytical expression is found for circumbinary orbits
with high accuracy using the expressions \ref{rpsolution},
\ref{psipsolution}, and the equations (\ref{eq:expansion_rho2},
\ref{eq:expansion_psip2}, \ref{eq:mu}, \ref{eq:disturbed_r} and
\ref{eq:disturbed_v}).

\subsection{The Radial Extension of Circumbinary Discs (The {\it Gap})}
\label{sec:size2}

Our purpose in this section is to provide a good estimation for the
radius of the inner boundary of a circumbinary disc. Then we are
looking for the closest stable orbits to the binary system. In this
case, it is required to calculate orbits decreasing in size until
consecutive orbits intersect each other. The procedure we follow was
to find a few new terms in the approximate solution for the disturbed
orbit and calculate the larger inner orbit that intersects a
contiguous one with the method described in Section \ref{sec:size1}.
In this case, as in the circumstellar disc calculation (see Section
\ref{sec:size1}), the angular correction $\psi_{p_i}$ is proportional
to $\sin\tau$, the intersections occur in the same manner at $\tau=0$
and $\tau=\pi$, i.e., in the line joining the stars. The larger the
number of terms considered for the approximation, the closer the orbit
is to the orbit calculated in PSA05. However, the difference in sizes
is large in spite of taking into account the terms up to
$\mu^{21/3}=({a_{np}\over a})^{-21/2}$ that gives high precision in the
circumstellar discs case. Table 2 gives for a set of values for
$M_{1}$, the non-perturbed radius $a_{np}$, the physical radius of the
orbit in the intersection with the line connecting the stars,
$r(\tau=0)$ and $r(\tau=\pi)$, and the analogous values in
PSA05. Here, for each $M_{1}$ there is an intersection at $\tau=0$ or
at $\tau=\pi$. Note that, for example, the system with $M_{1}=0.2$ is
the same system with $M_{1}=0.8$, only rotated an angle $\pi$, then,
the intersection at $a_{np}=1.3920$ in $\tau=0$ in the former one,
corresponds to $a_{np}=1.3920$ in $\tau=\pi$ in the latter. Thus, Table
2 gives radii for $M_{1}\leq 0.5$.

\begin{table*}
 \centering
% \begin{minipage}{140mm}
  \caption{Size of the central gap in the circumbinary disc}
  \begin{tabular}{@{}cccccc@{}}
%    Name     &            & \multicolumn{4}{c}{Flux density (Jy)%
   $M_{1}$ & $a_{np,\tau=0}$ & $r(\tau=0)$ & $r(\tau=\pi)$ & $r(\tau=0)_{PSA}
    $ & $r(\tau=\pi)_{PSA}$ \\
%        &  &  &  &  &  & group & (d) & curve \\
%        &  &  &  &  &  &       &     & type  \\[10pt]
 0.1 & 1.3547 & 1.5660 & 1.2956 & 1.87 & 1.80 \\
 0.2 & 1.3920 & 1.6267 & 1.3509 & 2.04 & 2.00 \\
 0.3 & 1.3733 & 1.6179 & 1.3637 & 1.94 & 1.90 \\
 0.4 & 1.3131 & 1.5604 & 1.3642 & 1.92 & 1.90 \\
 0.5 & 1.1643 & 1.4183 & 1.4183 & 2.00 & 2.00
\end{tabular}
%\end{minipage}
\end{table*}

\subsection{Stability Analysis}
\label{sec:stability}

In the numerical approach, the solution for a closed orbit is searched
by successive iterations. This means that an unstable orbit
(surrounded by orbits far from the solution) is very hard to find.
Unlike numerical calculations, analytically, one can calculate either
stable and unstable orbits without any possible identification. One
has to apply another criteria to look for the long-lived orbits as the
ones in real discs, this criteria is taken from a stability analysis.

In the case of circumstellar discs, the criteria to end up a disc was
the intersection of orbits. In the circumbinary discs case, this is not
applicable since in general orbits start being unstable before any
intersection of orbits.

Message (1959) defines an orbit as stable or unstable using the
equation that describes normal displacements from a periodic orbit in
the three-body restricted problem. These displacements are solved with
terms proportional to $\exp({ic\tau})$, where $c$ is given by

\begin{equation}
    c^{4}-(4+\lambda_{-1}+\lambda_{1})c^{2}+2(\lambda_{1}-\lambda_{-1})c+
          \lambda_{-1}\lambda_{1}-\nu_{1}\nu_{-1}=0,
    \label{eq:c}
\end{equation}

where

\begin{eqnarray}
    \lambda_{\pm 1} &=& \theta_{p_0}-1-(\beta_{0}+\beta_{\pm 2})
           \theta_{p_1}\theta_{p_-1}
         -\beta_{\pm 2}^{2}\beta_{\pm 3}\theta_{p_1}^{2}\theta_{p_-1}^{2}, \\
    \nu_{\pm 1} &=& -\beta_{0}\theta_{p_\pm 1}^{2}, \\
    \beta_{k} &=& {1\over \theta_{p_0}-(k+c)^2},
\end{eqnarray}

\noindent and $\theta_{p_0}$,$\theta_{p_\pm 1}$ are the main coefficients
of the Fourier expansion of $\Theta(\tau)$, where $\Theta(\tau)$ is
the function involved in the normal equation displacement,

\begin{equation}
    {d^{2}q\over d\tau^{2}}+\Theta(\tau)q=0,
\end{equation}

\noindent which is given in Message (1959). The function
$\Theta(\tau)$ depends on the shape of the orbit, i.e. depends on
$\tau$.If for a specific orbit, $c$ has an imaginary term then the
orbit is unstable. Calculation of $c$ is made for analytical orbits
given by equations (\ref{eq:disturbed_r} and \ref{eq:disturbed_v})
with $r_p$ and $\psi_p$ estimated with equations
(\ref{eq:expansion_rho2}, \ref{eq:expansion_psip2}) and the
coefficients given by equations \ref{rpsolution} and
\ref{psipsolution}. Orbits smaller and larger that the
intersecting-orbits given in Table 2 are considered. The values taken
for the parameter $a_{np}$ are given in Table 3.

\begin{table*}
 \centering
% \begin{minipage}{140mm}
  \caption{Values of the parameter $a_{np}$ for the orbits studied in the 
           instability analysis.}
  \begin{tabular}{@{}ccccccc@{}}
%    Name     &            & \multicolumn{4}{c}{Flux density (Jy)%
   $M_{1}$ & $a_{np,1}$ & $a_{np,Inters}$ & $a_{np,2}$ & $a_{np,3}$
    & $a_{np,PSA}$ & $a_{np,4}$ \\
%        &  &  &  &  &  & group & (d) & curve \\
%        &  &  &  &  &  &       &     & type  \\[10pt]
 0.1 & 1.25 & 1.3547 & 1.45 & 1.65 & 1.80 & 2.75 \\
 0.2 & 1.29 & 1.3920 & 1.49 & 1.69 & 1.99 & 2.79 \\
 0.3 & 1.27 & 1.3733 & 1.47 & 1.67 & 1.87 & 2.77 \\
 0.4 & 1.21 & 1.3131 & 1.41 & 1.61 & 1.86 & 2.71 \\
 0.5 & 1.06 & 1.1643 & 1.26 & 1.46 & 1.96 & 2.56
\end{tabular}
%\end{minipage}
\end{table*}

The parameter $a=a_{np,PSA}$ corresponds to the analytical orbit closer
to the innermost orbit of the circumbinary disc, found in PSA05, for
several stellar mass values, $M_{1}$. Thus, if this set of orbits is
(un)stable we expect that the numerical orbits also are (un)stable. We
solve equation (\ref{eq:c}) for the orbits in Table 3 and the results
are given in Table 4. There, the largest value of the imaginary part
of $c$, looking at all the modes found, are given by $M_{1}$, ranging
from $0.1$ to $0.5$.

\begin{table*}
\centering
% \begin{minipage}{140mm}
\caption{Maximum imaginary Part of $c$ for the Orbits Studied in the 
          Instability Analysis}
\begin{tabular}{@{}cccccc@{}}
%    Name     &            & \multicolumn{4}{c}{Flux density (Jy)%
   $Max|Im(c)|$ $\backslash$ $M_{1}$ & 0.1 & 0.2 & 0.3 & 0.4 & 0.5 \\
%        &  &  &  &  &  & group & (d) & curve \\
%        &  &  &  &  &  &       &     & type  \\[10pt]
 $c(a_{np,1})$ & $2.65\times 10^{-3}$ & $5.64\times 10^{-3}$ & $8.26\times 10^{-3}$
    &  $9.38\times 10^{-3}$ & $3.60\times 10^{-11}$  \\
 $c(a_{np,Inters})$ & $6.47\times 10^{-4}$ & $1.43\times 10^{-3}$ & $1.84\times 10^{-3}$
    &  $1.62\times 10^{-3}$ & $2.76\times 10^{-12}$   \\
 $c(a_{np,2})$ & $2.35\times 10^{-4}$ & $4.87\times 10^{-4}$ & $5.83\times 10^{-4}$
    &  $4.29\times 10^{-4}$ & $4.35\times 10^{-13}$ \\
 $c(a_{np,3})$ & $4.96\times 10^{-5}$ & $9.24\times 10^{-5}$ & $9.60\times 10^{-5}$
    &  $5.53\times 10^{-5}$ & $2.68\times 10^{-14}$ \\
 $c(a_{np,PSA})$ & $2.09\times 10^{-5}$ & $2.07\times 10^{-5}$ & $2.72\times 10^{-5}$
    &  $9.87\times 10^{-6}$ & $0$  \\
 $c(a_{np,4})$ & $0$ & $0$ & $0$ & $0$ &  $0$
\end{tabular}
%\end{minipage}
\end{table*}

From Table 4, the trend for $Max|Im(c)|$ is to decrease when the orbit
is moving away from the binary system, as expected. Note that
$Max|Im(c(a_{np,PSA}))|$ is quite small, consequently, this value can
be taken as a threshold for instability and the orbit in PSA05 will
naturally define the boundary between the unstable (not possible for
gas orbits) and the stable (possible) part of the disc.

It is worth noticing that this analysis can be applied either to
numerical or analytical orbits. Only the Fourier coefficients of a
function that depends on the orbit are required. Thus, the stability
analysis described here is general, and can be used with any orbit of
the restricted circular three-body problem.

\section{Test: Comparison with Numerical Results}\label{sec:comparison}
In this section we present a simple comparison between the results
presented in this work and precise numerical results performed with
the usual methods to integrate the motion equations in a circular
binary potential. We show for this purpose two tests, the first is a
direct qualitative comparison between numerical and the analytical
approximation to some representative discs. The second including
velocities, are the correspondent rotation curves.

\subsection{The Numerical Method}\label{numericalmethod}
The equations of motion of a test particle are solved for the circular
restricted three body problem using an Adams integrator (from the NAG
FORTRAN library). Cartesian coordinates are employed, with origin at
the center of mass of the binary, in an inertial reference frame. Here
we look for the families of circumstellar and circumbinary orbits at a
given stellar phase of the stars (equivalent to look for the families
in the non-inertial frame of reference). We calculate the Jacobi
energy of test particles, in the non-inertial frame of reference
co-moving with the stars, as a diagnostic for the quality of the
numerical integration, conserved within one part in $10^{9}$ per
binary period.

\subsection{Geometrical Comparison of Discs}\label{compvisual}
In the case of binary systems is well known the shape and extension of
discs are different from the single stars case, where the periodic
orbits are circles with a radial unlimited (by the potential of the
single star) extension. In binary systems the stellar companion exerts
forces able to open gaps, limiting the extension of the discs and
producing at the same time deformations in the shape of the discs
making them visually different from single star cases. In this section
we show a comparison between the discs obtained by the analytical
approximation provided in this work, the numerical solution and the
keplerian approximation.

To construct the analytical circumstellar discs we show in this
section, we have employed the equations \ref{eq:disturbed_r} and
\ref{eq:disturbed_v} that can be written in the inertial reference
system,

\begin{equation}
    r=a_{np}(1+r_p),
    \label{eq:disturbed_r2}
\end{equation}

\begin{equation}
    \psi=\tau+\psi_p,
    \label{eq:disturbed_v2}
\end{equation}

\noindent with $\tau$ in the interval $[0,2\pi]$.

\begin{figure}
\plotone{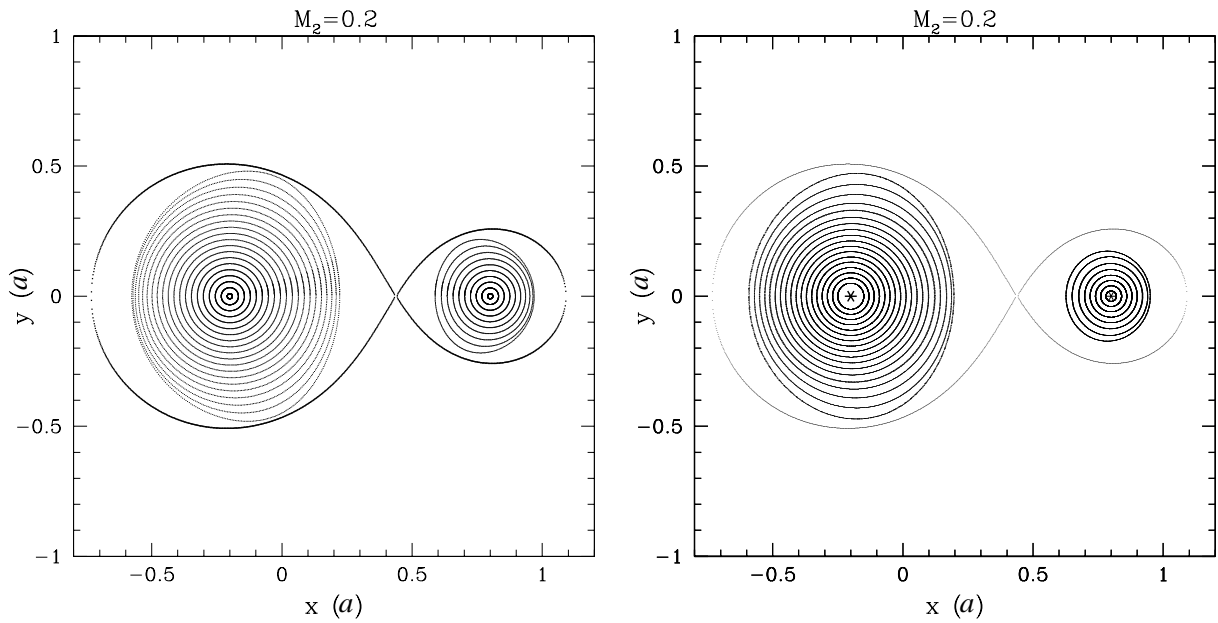}
\caption {Circumstellar discs comparison between orbits with the
  approximation presented in this work and numerical results described
  in subsection \ref{numericalmethod} for the case $M_2=0.2$. Left
  panel shows the analytical approximation provided in this work.
  Right panel shows the numerical approximation. The darker curve
  surrounding the discs is the Roche Lobe. The axes are in units of
  the distance between the stars, $a$}
\label{fig.compEB_mosaico_q0.2}
\end{figure}

In the Figure \ref{fig.compEB_mosaico_q0.2}, we show the comparison
for circumstellar discs between the approximation presented in this
work for periodic orbits in a binary system (left panels) and
numerical calculations (right panels). The value for $M_2$ is
indicated on the top of the figure. The darker curve represents the
Roche lobe.

\begin{figure}
\plotone{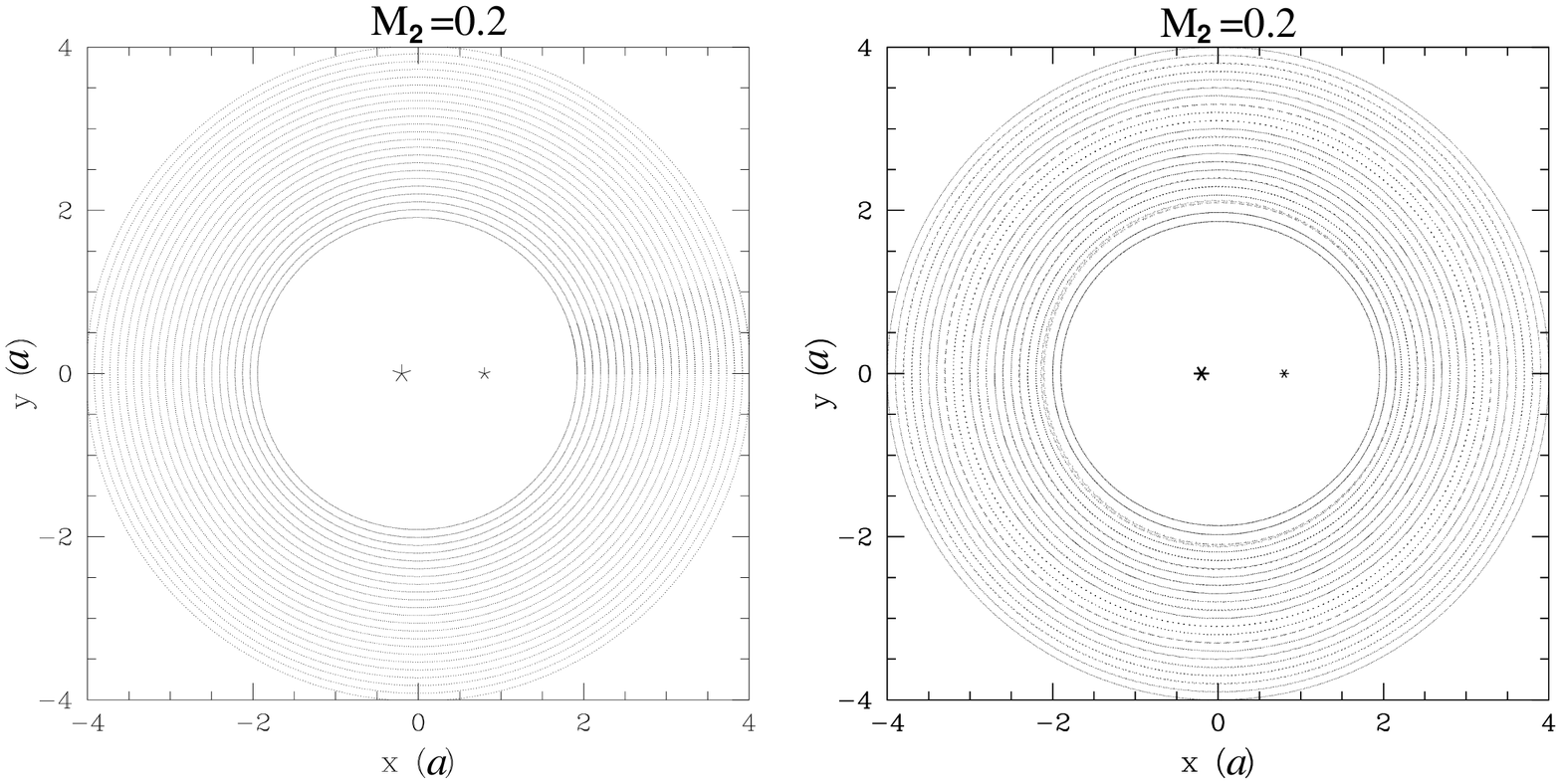}
\caption {Circumbinary discs comparison between orbits with the
approximation presented in this work and numerical results described
in subsection \ref{numericalmethod} for the case $M_2=0.2$. Left panel
shows the analytical approximation provided in this work. Right panel
shows the numerical approximation. The axes are in units of the
distance between the stars, $a$. The distance of the primary and
secondary star with respect to the center of mass of the system is
indicated by the black asterisks}
\label{fig.compEB_mosaico_q0.2_CB}
\end{figure}

Likewise, in Figure \ref{fig.compEB_mosaico_q0.2_CB} we show a
qualitative comparison of a circumbinary disc between the analytical
approximation (left panels) and numerical calculations (right
panels). The distance of the stars with respect the center of mass of
the binary is shown.

The discs present a slight deformation of the orbits depending on the
radius: the outermost orbits are farer from being circles. The orbits
are, a very good approximation, ellipses with an eccentricity
depending on the radius to the central star, in the cincumstellar
discs, and to the center of mass in the circumbinary discs case. Thus,
we have calculated the m=2 Fourier mode for all the orbits given by

\begin{eqnarray}
A_k=\frac{1}{N} \sum_{i=0}^{N} s(\phi_i) cos(k\phi_i),  \,
\nonumber\\ 
B_k=\frac{1}{N} \sum_{i=0}^{N} s(\phi_i) sin(k\phi_i) \, 
\label{LopEll},
\end {eqnarray}

\noindent where the index $i$ refers to the number of evenly
distributed (by interpolation) in angle points in a given loop, and
$N$ is the total number of points in the loop, $s(\phi_i)$ is the
distance to the point $i$ from the center of mass of the binary.  In
this manner, the average inner radius of a circumbinary disk is given
in units of the semimajor axis $a$ by the coefficient $A_K$ with
$k=0$. The ellipticity ($ell$) by $\sqrt{A_k^2+B_k^2}$ with $k=2$ in
the same units and transformed first to the ratio $b_i/a_i$, where
$a_i,b_i$ are the semimajor and semiminor axis of the orbit $i$, and
finally transformed to eccentricity ($ecc$) by,

\begin{equation}
ecc=\sqrt{1-ell^2}
\end {equation}

\noindent which represents a more sensitive geometrical characteristic
of the orbits than ellipticity. We have not considered in this
comparison higher Fourier modes since they are negligible compared to
the m=2 mode.

\begin{figure}
\plotone{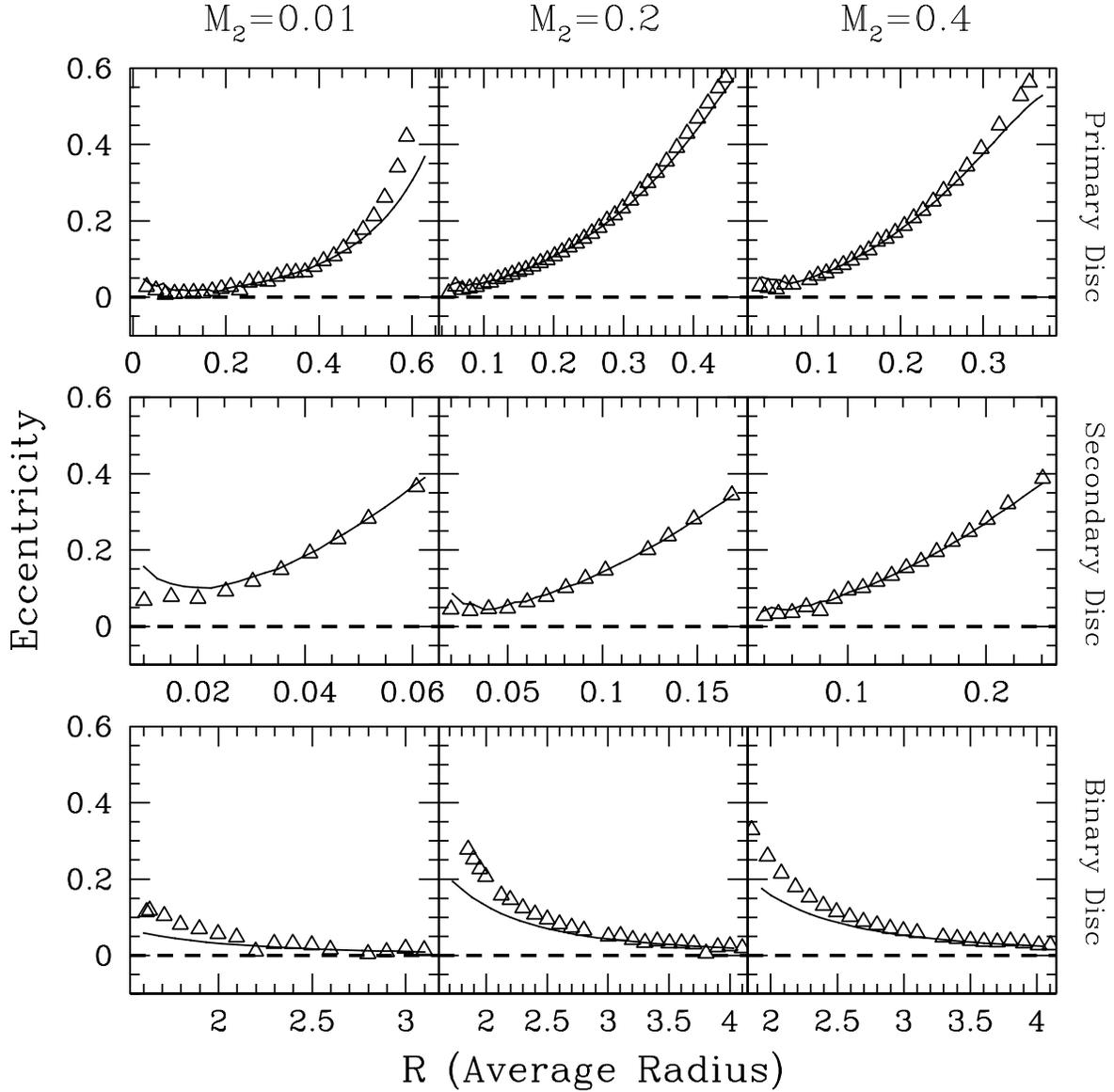}
\caption {Comparison between the orbital eccentricity vs radius
produced by the analytical approximation provided in this work
(continuous lines), the full numerical solution (open triangles), and
for reference the keplerian ($ecc=0$) solution (dashed straight
lines). Three examples of different masses, $M_2= 0.01, 0.2, 0.4$, are
indicated by the upper labels at left, middle and right panels
respectively. The upper frames are referred to the circumprimary discs,
the middle to the circumsecondary discs, and the lower are the
circumbinary discs}
\label{fig.mosaic_ecc}
\end{figure}

In the Figure \ref{fig.mosaic_ecc} we show the eccentricity vs the
average radius for a) the orbits produced by the analytical
approximation provided in this work (continuous lines in the 9
frames), b) the full numerical solution (open triangles), and for
reference c) the keplerian ($ecc=0$) solution (dashed lines). In this
figure we show three different masses $M_2= 0.01, 0.2, 0.4$, at left,
middle and right panels respectively (as indicated on the upper frames
labels). The upper frames are referred to the circumprimary discs, the
middle to the circumsecondary discs, and the lower are the
circumbinary discs. The system of reference for each case is on the
respective star for the circumstellar discs and in the center of mass
for the circumbinary discs.

\subsection{Rotation Curves}\label{compcuantitativa}

We also show here, for a quantitative comparison involving positions
and velocities, the rotation curves for binary systems with
$M_2=0.01,0.2,0.4$.

We have constructed the rotation curves by direct differentiation of
equations \ref{eq:disturbed_r} and \ref{eq:disturbed_v} (or more
precisely of equations \ref{eq:disturbed_r2} and
\ref{eq:disturbed_v2}), to obtain the total velocity ($v_c$) at two
different regions of the discs. For the primary discs we compared the
rotation curves obtained along the x axis, on the left part of the
discs (where the radial velocities are negative and prograde with the
rotation of the binary). For the secondary we chose to compare the
rotation curves along the x axis but on the right part of the disc
(where the radial velocities are positive and prograde with the
rotation of the binary). Once the velocities are calculated, we
transformed the results to the inertial frame (in the circumstellar
discs) by simply adding (or subtracting depending on the given disc)
the velocity of the correspondent star, and adding or subtracting also
a factor $\Omega r$, where $\Omega$ is the angular velocity of the
star and $r$ is the radius of a given point in he rotation curve. The
last because of the chosen reference system where this work solves the
equations, which anchorages the discs to rotate with the system.

\begin{figure}
\epsscale{0.7}
\plotone{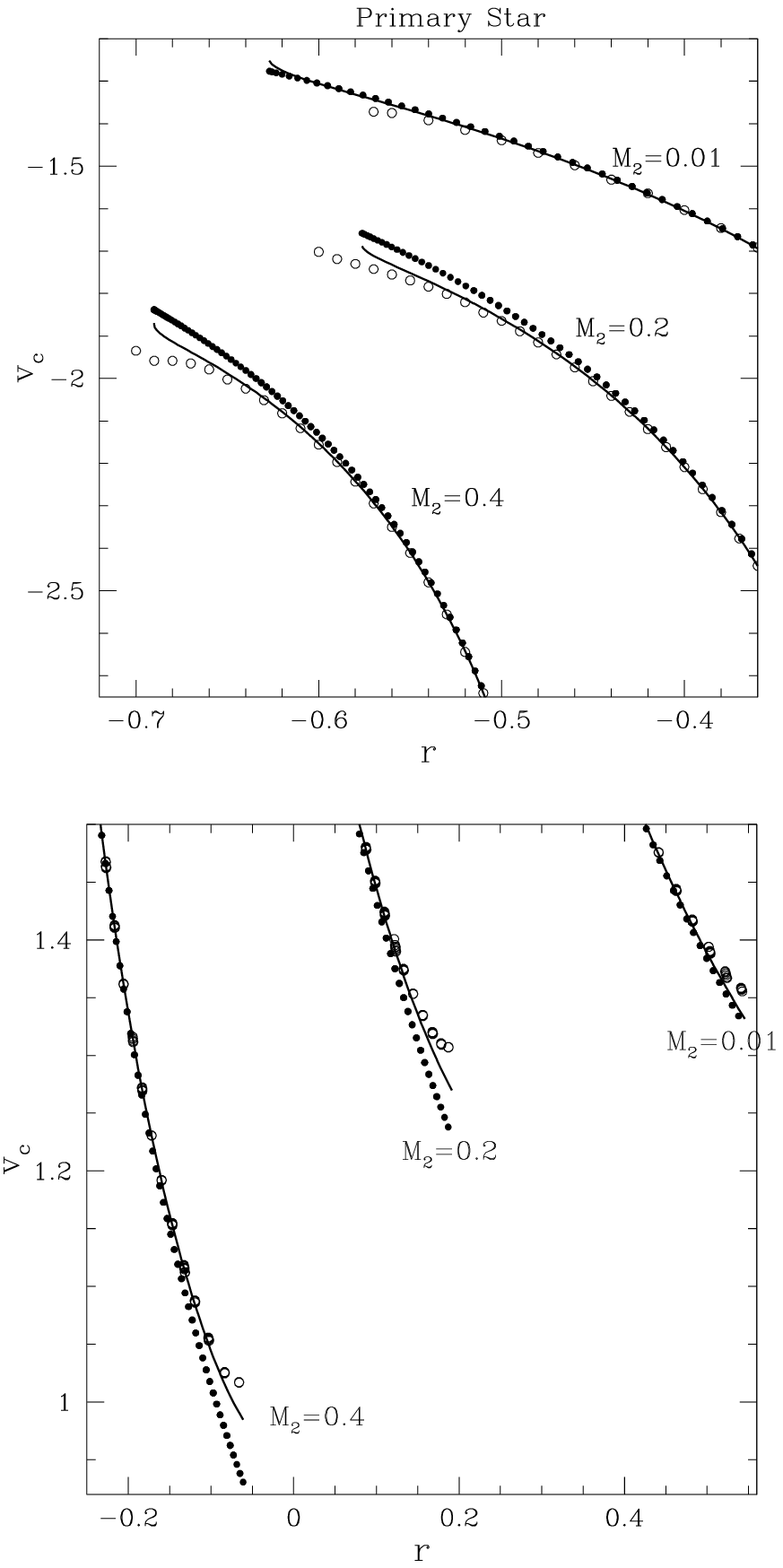}
\caption {Comparison between the keplerian rotation curve (filled
circles), the analytical approximation provided in this work
(continuous line), and the numerical solution (open circles), for the
outer region (top frame), and inner region (bottom frame) of the
primary disc, in the different cases $M_2=0.01,0.2,0.4$.}
\label{fig.crot_s1}
\end{figure}

\begin{figure}
\epsscale{0.7}
\plotone{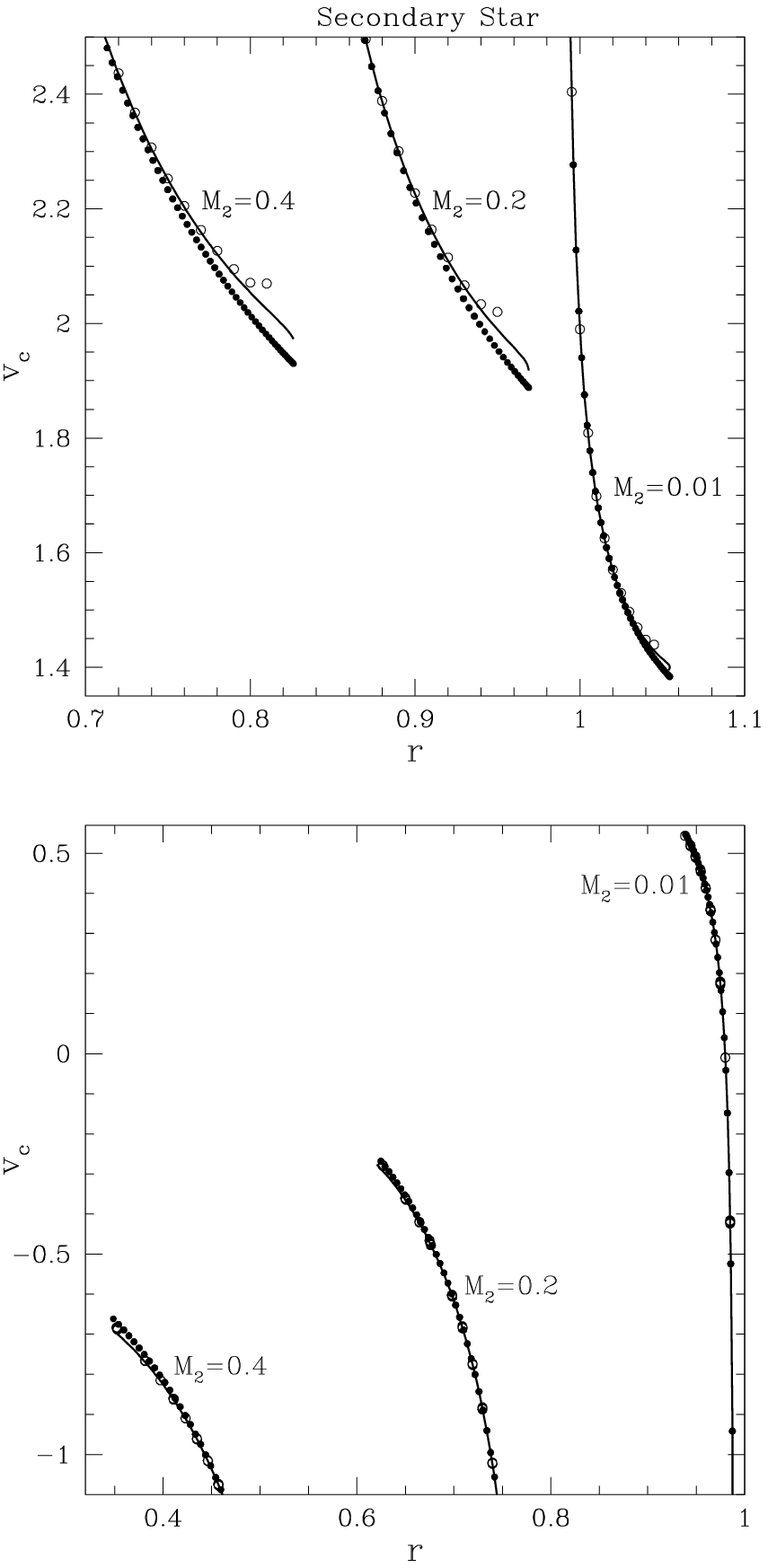}
\caption {Same as Figure \ref{fig.crot_s1} but for the secondary star}
\label{fig.crot_s2}
\end{figure}

In the Figure \ref{fig.crot_s1} and \ref{fig.crot_s2} we show the
outer (top panels) and inner (bottom panels) regions of the primary
and secondary rotation curve discs, respectively. By ``outer'' or
``inner'' regions we mean, in a reference system located in the
primary (or in the secondary) star, measuring and angle from the line
that joins the stars, the inner region corresponds to angle zero from
this line, and the outer regions correspond to and angle of
$180^o$. For both figures we plot the three different cases,
$M_2=0.01,0.2,0.4$. Three different types of lines indicate, a)the
keplerian rotation curve on the top (filled circles); b) the
analytical approximation provided in this work (continuous line); c)
the numerical solution (open circles).  The velocity and radius are
given in code units (in such a way that $G=1$, $M_1+M_2=1$, $a=1$ and
$\Omega=1$).

The rotation curves are in approximately 70$\%$ of the total radius of
the correspondent disc, keplerian curves. For the last 30$\%$ the
keplerian discs velocity are systematically over the numerical result
that solves with high precision the restricted three body problem. The
analytical solution we provide here is very close to the numerical
solution as expected.

\begin{figure}
\epsscale{0.7}
\plotone{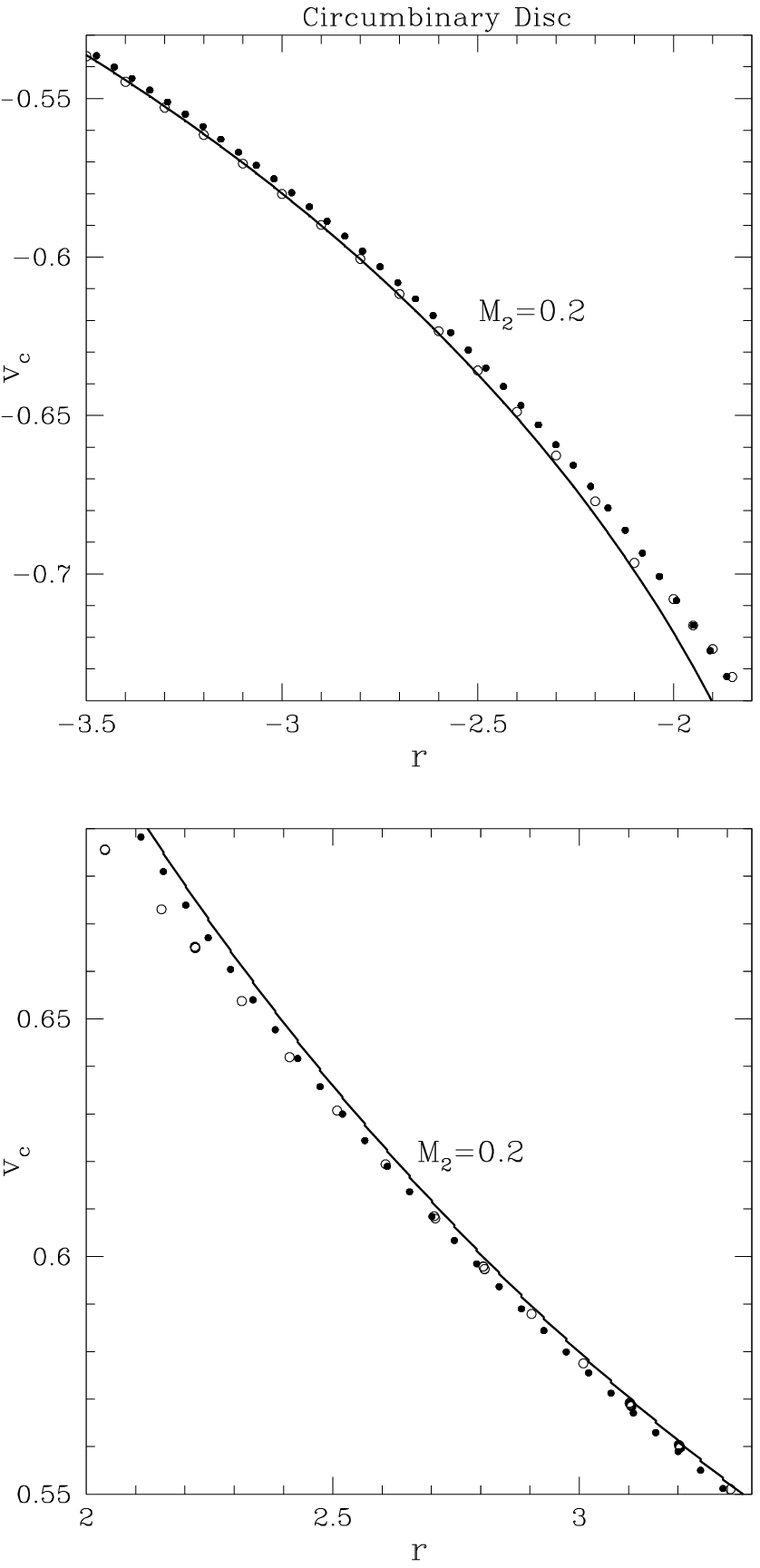}
\caption {Comparison between the keplerian rotation curve (filled
circles), the analytical approximation provided in this work
(continuous line), and the numerical solution (open circles), for the
left region (top panel), and right region (bottom panel) of the
circumbinary disc, for $M_2=0.2$.}
\label{fig.CR_CB}
\end{figure}

In the Figure \ref{fig.CR_CB} we show the rotation curves as in
Figures \ref{fig.crot_s1} and \ref{fig.crot_s2}, but for the
circumbinary disc. Although in this case both approximations (the one
given in this work, and the numerical one) are very close to a
keplerian rotation curve, it is not exactly keplerian. The velocities
in the analytical approximation are sistematically under the keplerian
curve and they are practically the same as the ones provided by the
numerical solution, until the end (the beginning of the gap) where the
numerical solution goes slightly over the analytical solution. We
present only one case of mass ratio because other mass ratio would
give very similar results.

In the case of the full (numerical) solution, it is worth to notice we
obtain in some cases discotinuities on the rotation curves and orbits
mainly due to resonances. This is the case, for instance, in the
Figure \ref{fig.CR_CB}, where we appreciate a discontinuity very close
to the resonance 3:1 ($r\sim -2.1 a$ in the figure). The discontinuity
is also noticeable in the Fig. \ref{fig.compEB_mosaico_q0.2_CB} at the
same radius. Due to the perturative fit performed in this work,
finding this kind of discontinuities due to resonances, that result in
general very narrow in radius but where the physics is highly
non-linear is out of the reach of our approach.

\section{CONCLUSIONS}\label{sec:conclusions}

We have constructed an analytical set of equations based on
perturbative analysis that result simple and precise to approximate
the solution for the circular restricted three body problem. We choose
some terms in the expanded equations of motion, which are relevant to
the solution of the problem. The original equations are approximated
with a set of equations that, in many cases, have an analytical
solution. We show the goodness of our analytical approximation by
direct comparison of geometry and rotation curves with the full
numerical solution. 

The set of equations we prsent can provide any periodic orbit for a
binary system, either circumstellar or circumbinary, and not only the
outer edge radius of circumstellar discs or the inner edge radius
(gap) of the circumbinary disc, without the need to solve the problem
numerically.

The relations provided are simple and straightforward in such a way
that our approximation can be used for any application where initial
conditions of periodic orbits or complete periodic orbits are needed,
or for direct study of the three body restricted circular problem. For
instance, in order to introduce on a hydrodynamical or particle code,
our initial conditions would result in much more stable discs than
with keplerian initial conditions, or than by constructing the discs
directly from hydrodynamical simulations of accretion to binaries that
will require long times to obtain stable discs. Since our
approximation is completely analytical it will have the obvious
additional advantage of being computationally, extremely cheap, and
easier to implement than the numerical solution.

The periodic orbits respond uniquely to the binary potential and do
not consider other physical factors such as gas pressure, or viscosity
that work to build the fine details of discs structure. They are
however, the backbone of any potential and their shape and behavior
give the general discs phase space structure. In this manner, apart of
all the possible hydrodynamical or particle discs applications, we can
directly use them to study from the general discs geometry to rotation
curves, or rarification and compression zones by orbital crowding,
etc.

\section*{Acknowledgments} 
We acknowledge Linda Sparke, Antonio Peimbert and Jorge Cant\'o for
enlightening discussions. B.P. thanks project CONACyT, Mexico, through
grant 50720.

\end{document}